\documentclass[twocolumn,showpacs,amsmath,amssymb,prb,citeautoscript]{revtex4}
\usepackage{graphicx}
\usepackage{dcolumn}
\usepackage{bm}
\usepackage{times}
\begin{document}

\title{Cluster Algorithms for Quantum Impurity Models and
       Mesoscopic Kondo Physics}
\author{Jaebeom Yoo}
\author{Shailesh Chandrasekharan}
\author{Ribhu K.~Kaul}
\author{Denis Ullmo}
\thanks{Permanent address: Laboratoire de Physique Th\'eorique et
  Mod\`eles Statistiques
(LPTMS), 91405 Orsay Cedex, France.}
\author{Harold U.~Baranger}
\affiliation{Department of Physics, Duke University, Durham, NC 27708-0305}

\date{\today}

\begin{abstract}
Nanoscale physics and dynamical mean field theory have both generated increased interest in complex quantum impurity problems and so have focused attention on
the need for flexible quantum impurity solvers. Here we demonstrate that the mapping of single quantum impurity problems onto spin-chains can be exploited to yield a powerful and extremely flexible impurity solver. We implement this cluster algorithm explicitly for the Anderson and Kondo Hamiltonians, and illustrate its use in the ``mesoscopic Kondo problem''. To study universal Kondo physics, a large ratio between the effective bandwidth $D_\mathrm{eff}$ and the temperature $T$ is required; our cluster algorithm treats the mesoscopic fluctuations exactly while being able to approach the large $D_\mathrm{eff}/T$ limit with ease. We emphasize that the flexibility of our method allows it to tackle a wide variety of quantum impurity problems; thus, it may also be relevant to the dynamical mean field theory of lattice problems.
\end{abstract}


\maketitle

The Kondo problem, which describes the coupling of a quantum magnetic impurity to an electron gas, has a rich history in condensed matter physics \cite{HewsonBook,Kondo64,Wilson75,Andrei80,Wiegmann80,HirFye86,FyeHir88,FyeHir89}.  Recent experimental and theoretical developments have generated considerable renewed interest in this field. On the experimental side, nanostructures in ultra-small metal particles, semiconductor heterostructures, carbon nanotubes, and organometallic molecules all show physics that can be mapped onto quantum impurity problems \cite{Lee88,Glazman88,Goldhaber98,Gold98,Cronenwett98,VanDerWeil00,NygCobLin,ParPasJon,LiaShoBoc}. On the theoretical side, dynamical mean field theory (DMFT) has reduced lattice problems in infinite dimensions to self-consistent quantum impurity problems \cite{GeoKot}. Nanoscale physics and DMFT have furthermore emphasized the richness and variety of these questions: In mesoscopic physics quantum dots behave as impurities; their tunability, as well as the new energy scales implied by the confined geometry, have lead to several new impurity problems. In DMFT, lattice models map onto complex self-consistent quantum impurity problems that may not have arisen in previous contexts. Clearly, the paradigm of a spin-1/2 magnetic impurity in a metallic host proposed by Kondo\cite{Kondo64} is only a starting point in exploring a vast array of physically relevant quantum impurity problems.

In order to understand quantum impurity problems in depth and to compare with experiments, it is desirable to have exact results. Since exact analytic solutions \cite{Andrei80,Wiegmann80} make approximations that are invalid beyond the simplest situations, it is necessary to resort to numerical approaches. The need for precise numerical solvers has been especially highlighted by DMFT \cite{GeoKot}. Many numerical methods have been developed for quantum impurity problems, including the numerical renormalization group (NRG) \cite{Wilson75} and the quantum Monte Carlo algorithm introduced by Hirsch and Fye \cite{HirFye86}. Although these methods are powerful, they have individual weaknesses related to the principles on which they are formulated. Here we propose a new quantum Monte Carlo method for quantum impurity problems. Being formulated on a different principle from, for instance, either NRG or the Hirsch and Fye Monte Carlo method, this new approach has a different set of strengths and weaknesses; it is, in fact, more efficient than the traditional methods in some cases where these latter have difficulties.

One example where the weaknesses of the standard methods show up is the ``mesoscopic Kondo problem''.  Nanoscale experiments have motivated new questions that are usually ignored in the study of quantum impurity problems. One may ask, for instance, what is the effect of confining the bath electrons to a finite, fully coherent, region of space. In this case, the local electronic density of states fluctuates strongly, and a flat band description is no longer applicable.  In addition to the effects caused by a finite mean level spacing $\Delta$, \cite{Thimm99,Simon02,Cornaglia02} confinement leads to mesoscopic fluctuations associated with interference effects \cite{Kaul03,Kaul05} that cannot be captured exactly by analytic methods.  Numerically, NRG is not natural for a situation with discrete levels, and involves some approximations (essentially neglecting part of the conduction electrons' Hilbert space) that have yet to be tested for structured density of states.  Although the Hirsch and Fye algorithm does not depend on the conduction band, the formulation of the algorithm makes it impractical to use in the most interesting regime when $D_{\rm eff}$ (the effective bandwidth where the impurity is magnetic) becomes large compared to the temperature $T$. Hence, a detailed study of the effects of mesoscopic fluctuations on universal Kondo physics requires new numerical tools, able to handle any kind of discrete density of states while allowing the study of large $D_\mathrm{eff}/T$. The goal of this paper is to show that such a QMC algorithm indeed exists.

We develop our QMC method explicitly for two quantum impurity problems, the Anderson and Kondo models, and point out later that our algorithm is very flexible and has a broad range of applications. The Hamiltonian for both models can be written as a sum of two terms $H  \!=\!  H_0 + H_1$ where $H_0$ describes the free electron band, $H_0  \!=\!  \sum_{\alpha= 1}^N \sum_{\sigma \pm 1} \epsilon_\alpha c^\dagger_{\alpha\sigma} c_{\alpha\sigma}$, and is entirely specified by the energies $\epsilon_\alpha$ and the eigenfunctions $\phi_\alpha({\bf r})$ of the confined electron gas.  We will assume $-D \!\leq\! \epsilon_\alpha \!\leq\! D$, where $D$ is the bandwidth. A ``flat'' band would correspond to the choice $\epsilon_\alpha  \!=\!  \alpha \Delta$ and $|\phi_\alpha({\bf r})|^2  \!=\!  1/\Omega$, which, for temperatures larger than the mean level spacing $\Delta$ would just reproduce the bulk behavior.  Here $\Omega$ is the volume of the electron gas. For fully coherent confined systems, however, both of these quantities display mesoscopic fluctuations from level to level, or as a function of some external parameter.  For chaotic systems with time reversal symmetry, a good model for these fluctuations is provided by the Gaussian Orthogonal Ensemble (GOE) of random matrix theory \cite{Bohigas91}.  The averaged quantities are kept the same as in the flat band case, and in particular $\langle \rho(\epsilon) \rangle  \!=\!  \rho_0  \!=\!  1/(\Delta\Omega)$, where $\langle \rho(\epsilon)\rangle$ is the average over realizations of the local density of states $\rho(\epsilon)  \!=\!  \sum_\alpha |\phi_\alpha({\bf r})|^2 \delta(\epsilon-\epsilon_\alpha)$. In the following, we shall study both a flat band and some realizations drawn from GOE for illustration.

The interactions of the conduction electrons with the impurity are
encoded in $H_1$ along with the impurity Hamiltonian. For the Anderson
model \cite{Anderson61}
\begin{eqnarray} 
H_1^\textrm{A} & = & 
  V \sum_\sigma  \left[\Psi^\dagger_\sigma({\bf r_0}) d_\sigma  
+  d^\dagger_\sigma \Psi_\sigma({\bf r_0}) \right]
   \\ \nonumber
&+& \sum_\sigma \epsilon_d d^\dagger_\sigma d_\sigma 
+ U d_\uparrow^\dagger d_\uparrow d^\dagger_\downarrow
  d_\downarrow \, . \label{eq:Hi1}
\end{eqnarray}
where $\Psi^\dagger_\sigma({\bf r_0})  \!=\!  \sum_{\alpha} \phi_\alpha^*({\bf
r_0}) c^\dagger_{\alpha\sigma}$ creates an electron of spin $\sigma$ at the
location ${\bf r_0}$ of the impurity.  Similarly for the Kondo model
\cite{Wilson75}
\begin{equation}
H_1^\textrm{K} = \frac{J}{2} \sum_{\sigma \sigma'} \Psi^\dagger_\sigma({\bf r_0}) 
\vec{\bf \sigma}_{\sigma \sigma'} \Psi_{\sigma'}({\bf r_0}) 
\cdot \vec{{\bf S}^d} \; , \label{eq:Hi2}
\end{equation}
where $\vec{{\bf S}^d}$ is the impurity spin and $\vec{\bf \sigma}$ is the vector formed by the Pauli matrices. 
For a flat band and when $\epsilon_d = -U/2$ (symmetric case) the Anderson model turns into the Kondo model 
when $U\rightarrow \infty$ with $J\rho  \!=\!  8 \Gamma/\pi U$ and $D$ fixed. Here we have defined 
$\Gamma \!=\! \pi \rho_0 V^2$.

The basic strategy we are going to apply here relies on the fact that
the single-impurity Hamiltonians are essentially one-dimensional problems. 
Indeed, for both the Anderson and Kondo models, the impurity couples to 
the electron gas only locally, at ${\bf r_0}$.  Starting from 
$f_{1\sigma}  \!=\!  \Psi_\sigma({\bf r_0})$, one can find a basis of one 
particle states $f_{i\sigma},i \!=\! 1,2..,N$ such that the non-interacting
Hamiltonian $H_0$ becomes tridiagonal\cite{HewsonBook,Wilson75,Kri80,RaaUhr}:
\begin{equation}
H_0 = \sum_\sigma \sum_{i=1}^N \big(\alpha_i f_{i\sigma}^\dagger f_{i\sigma}
+\beta^*_i f_{i\sigma}^\dagger f_{i+1\sigma} + \beta_i
f_{i+1\sigma}^\dagger f_{i\sigma} \big) 
 \label{eq:Hchain}
\end{equation}
with $\beta_N \!=\! 0$. In this form we see that $H_0$ describes two open fermion chains. Each open fermion-chain is identical to a spin-chain since fermions hopping in one dimension cannot permute. Thus, there is no fermion sign problem to worry about when constructing quantum Monte Carlo algorithms.  Adding the interaction term 
$H_1$ merely couples the two fermion-chains at $i \!=\! 1$ with the impurity (see, for example, the illustration in Ref.\ \ \onlinecite{RaaUhr}).
In this form, one merely has to solve a spin-chain problem. Today spin-chain problems can be solved very efficiently in continuous time \cite{BeaWie,Eve} using the recently developed directed-loop cluster algorithm \cite{SylSan}. In the present case, to make the algorithm efficient one must perform two types of directed-loop updates: One that changes the fermion occupation numbers and the other that flips the fermion spins (changes $\sigma$). Since the directed-loop algorithm is well established, we will not discuss it here.

\begin{figure}
\includegraphics[width=0.9\hsize]{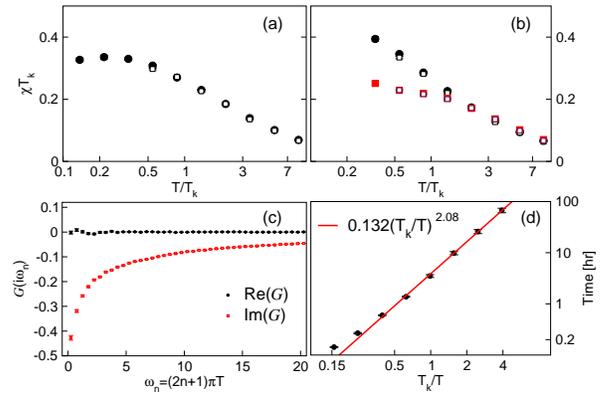}
\caption{\label{fig:HF} 
Comparison of the local susceptibilities $\chi$ obtained using the spin-chain algorithm (circles) and the Hirsch 
and Fye algorithm (open squares) in the symmetric Anderson model for $N \!=\! 5000$, $2D \!=\! 40$, and 
$U/\Gamma \!=\! 8/1.6$ ($T_{\rm K}\!=\!0.146$). $\chi T_\textrm{K}$ is in units of $(g\mu_\textrm{B}/2)^2$.
Three different realizations are shown: (a) Clean case and (b) two 
realizations drawn from GOE. 
(c) The real and imaginary parts of the thermal Green function 
versus Matsubara frequencies for the clean case at $T/T_\textrm{K}=0.55$.
(d) The time required in the spin-chain algorithm to obtain 
two percent errors on $\chi$ as a function $T_{\rm K}/T$.  In the low temperature regime, an almost 
quadratic dependence in $1/T$ is observed (compared to $1/T^3$ for a single sweep at fixed $\Delta \tau$ in 
the Hirsch and Fye algorithm\cite{HirFye86}). The value of $T_{\rm K}$ used was obtained from the two-loop 
RG equation of the Kondo model. 
} 
\end{figure}

We have implemented this algorithm for both the Anderson and the Kondo models using the continuous-time 
path integral directed-loop algorithm\cite{SylSan}.  Although the algorithm is equally applicable to 
the symmetric and the asymmetric Anderson models, here we focus only on the symmetric case for convenience.
The two directed-loop updates discussed above can readily give two correlation functions: the 
impurity one-particle thermal Green function $G_\textrm{d}(\tau)$ and the local susceptibility $\chi$.
$G_\textrm{d}(\tau)$ for the impurity can be measured while directed-loops for changing the fermion
occupation numbers are being constructed\cite{DorTro}. Moreover, the Green function at Matsubara 
frequencies, $G_\textrm{d}(i\omega_n) \!=\! \int_0^{1/T} \!\!d\tau\, \langle d(\tau) d^\dagger(0)\rangle e^{i\omega_n\tau}$,
necessary in a DMFT calculation can be directly measured since the Fourier integral involved in 
$G_\textrm{d}(i\omega_n)$ can be performed analytically during the loop construction in the 
continuous-time path integral implementation of the directed-loop algorithm. Similarly,
$\chi\!=\! \int_{0}^{1/T} \!\!d\tau\, \langle {S_z}^{d}(\tau){S_z}^{d}(0)\rangle$,
can be measured while directed-loops for flipping the fermion spins are being constructed.
We have tested our algorithm against exact diagonalization methods on small systems. As a further check, we have reproduced the results obtained from the Hirsch and Fye algorithm for the Anderson model. In Fig.~\ref{fig:HF} we compare results for three different realizations of $H_0$. We see that there is perfect agreement between the Hirsch and Fye and the spin-chain approach in the Anderson model. The thermal Green function for Matsubara frequencies is also shown for the clean case [Fig.~\ref{fig:HF}(c)].

\begin{figure}
\includegraphics[angle=270,width=0.9\hsize]{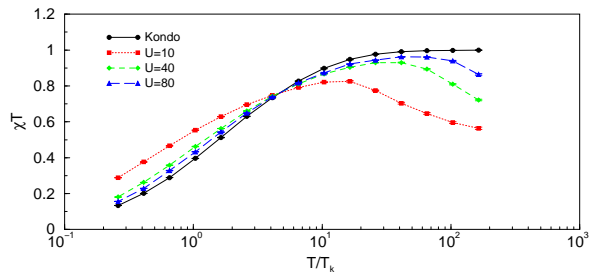}
\caption{\label{fig:AK} 
Local susceptibilities in the symmetric Anderson model
for various values of $U$ at fixed $8\Gamma/\pi U  \!=\!  J\rho  \!=\!  1/\pi$, $N \!=\! 1000$,
$2D \!=\! 10$ ($T_{\rm K}=0.122$) for a flat band. Data for the Kondo model at the same $J$ and $\rho$ is 
also shown. The value of $T_{\rm K}$ is obtained from the two loop RG equation in the Kondo model.
}
\end{figure}

To test our algorithm at larger $U$, we cannot compare to the Hirsch and Fye algorithm as it is very difficult to make $U$ large in that approach.
Hence, we compare our
results to those of a general algorithm we recently developed to study a variety 
of quantum impurity problems \cite{YooCha}. The local susceptibility obtained in the spin-chain algorithm
matches very well the result published in Ref.\ \ \onlinecite{YooCha} (parameters: 
$N \!=\! 2000$, $2D \!=\! 20$, $U \!=\! 25$, and $U/\Gamma \!=\! 4$). In Fig.~\ref{fig:AK} we 
show how the Anderson model results approach those of the Kondo model 
as $U$ becomes large at fixed $J$ \cite{FyeHir89,SchWol}.

We believe that the effort in the spin-chain cluster algorithm scales as a function of the system size as, in dimensionless units, either $(T_{\rm K}/T)(D/\Delta)$ or $(|\beta_1|/T)(D/\Delta)$, whichever dominates.
This shows that there is one disadvantage of the spin-chain approach: Our inability to deal with a continuum density of states, i.e. $\Delta \!=\! 0$. As one approaches this limit, the number of sites in the chain, $N$, will grow, and this in turn will increase the effort in updating the chain. As we have demonstrated, without any special effort we can simulate systems with $N \!=\! 5000$ on desktop computers; this could presumably be increased somewhat without too much difficulty. On the other hand, in the Hirsch and Fye approach the effort does not depend on $\Delta$. We can mitigate this disadvantage by noticing that in the spin-chain approach it is relatively straightforward to apply a logarithmic blocking of the energy levels very similar to the one introduced by Wilson \cite{Wilson75} in his work on the NRG.  Similar ideas have also been developed by several authors to study pseudo-gap problems \cite{CheJay,GonIng} in single-impurity models.

The starting point of the logarithmic blocking is to divide the
bandwidth $[-D,D]$ into a finite number of energy interval
$I_1,I_2...$ ad $I_{-1},I_{-2},...$ where $I_n
 \!=\! [D\Lambda^{-n},D\Lambda^{1-n}]$ for $n >0$ and $I_{n}
 \!=\! [-D\Lambda^{1+n},-D\Lambda^{n}]$ for $n < 0$, with a constant
$\Lambda>1$. As $|n|$ increases, the size of a logarithmic bins becomes
so small that there may contain only a few states.  When this number
is less than some constant $l$, we stop the logarithm discretization
and keep all the remaining states.  The optimal value for $l$ should
be determined by trial and error, but in the following we will just
take it to be one. For each interval $I_n$, we define the operator
\begin{equation}
a_{n\sigma} = \frac{1}{M_n}\sum_{\epsilon_\alpha\in I_n}\phi_\alpha({\bf r_0}) c_{\alpha\sigma}
\end{equation}
where $M_n  \!=\!  ( \sum_{\epsilon_\alpha\in I_n} \phi_\alpha^2)^{1/2}.$ Note that since $\Psi^\dagger_\sigma({\bf r_0})  \!=\!  \sum_n M_n a^\dagger_{n\sigma} $, the interaction terms in the Anderson and Kondo Hamiltonians depend only on $a_{n\sigma}$ and $a^\dagger_{n\sigma}$.  Thus, the only approximation in the blocking scheme comes from replacing $H_0$ by $\tilde H_0  \!=\!  \sum_n e_n a_{n}^\dagger a_{n}$ where $e_n  \!=\!  \sum_{\epsilon_\alpha \in I_n} \epsilon_\alpha |\phi_\alpha|^2/M_n^2$.  Note that as $\Lambda$ approaches $1$, $\tilde H_0 \rightarrow H_0$; the blocking disappears and all states are taken into account exactly.  Blocking reduces the the number of sites necessary in the spin-chain formulation to $N_\Lambda \sim \log N$. For a constant mean density of energy levels, $ N_\Lambda\simeq \log{N}/\log{\Lambda}$ which even for moderate $\Lambda$ represents a very significant reduction.

\begin{figure}
\begin{center}
\includegraphics[width=0.8\hsize]{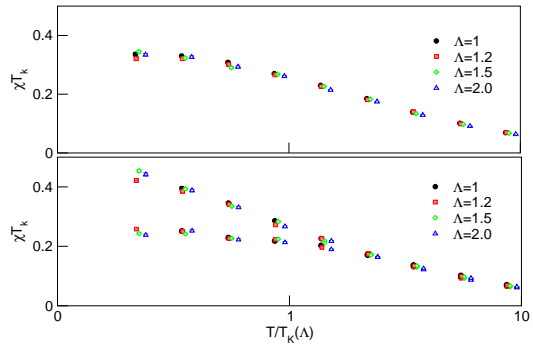}
\caption{\label{fig:Log} 
Local susceptibilities using logarithmic blocking, for the three realizations shown in 
Fig.~\ref{fig:HF} -- the clean case (top) and with mesoscopic fluctuations (bottom). The values 
of $N_\Lambda$ are roughly $80$ at $\Lambda \!=\! 1.2$, $40$ at 
$\Lambda \!=\! 1.4$, and $24$ at $\Lambda \!=\! 2$. The Kondo temperature $T_\textrm{K}(\Lambda)$ is 
obtained using the renormalized $J_\textrm{eff}(\Lambda)$ for logarithmic discretization:\cite{Kri80} $T_\textrm{K}$=0.145, 0.141, and 0.132, for $\Lambda$=1.2, $\Lambda$=1.5, and $\Lambda$=2.0, 
respectively.
}
\end{center}
\end{figure}

In order to check the usefulness of the logarithmic blocking we studied three different Hamiltonians (one clean and two realizations drawn from GOE). For each case we calculated the local susceptibilities using the blocked Hamiltonians with $\Lambda \!=\! 1.2$, $1.5$, and $2.0$. Our results are shown in Fig.~\ref{fig:Log}. One can see that the results from the blocked Hamiltonian approach the result without blocking ($\Lambda \!=\! 1$) as $\Lambda$ gets closer to $1$; all the $\Lambda \!=\! 1.2$ data are almost statistically indistinguishable from the exact results. Clearly, the optimal value of $\Lambda$ will depend on the parameters of the problem, the observables measured, and their needed accuracies. However, the reduction in the number of electron sites needed is dramatic even for small $\Lambda$. In the example here, one could reduce the number of sites from $5000$ to roughly $80$ with $\Lambda \!=\! 1.2$, to roughly $40$ with $\Lambda \!=\! 1.5$, and to roughly $24$ with $\Lambda \!=\! 2.0$.

The inability to treat a continuous density of state, even if it can be mitigated by the logarithmic blocking described above, will impose some limitations on the scope of application of our spin-chain cluster algorithm. It has, however, a number of strengths which we review briefly now, and compare more particularly to the well-established Hirsch and Fye method \cite{HirFye86}.
Some of the advantages of the new method are: 
(1) One remarkable feature of our algorithm is illustrated in Fig.~\ref{fig:HF}(d), which shows the computer time needed to obtain the local susceptibility within a fixed statistical error as a function of temperature. We find that the time grows approximately as $1/T^{2}$. For the Hirsch and Fye method the time for a QMC sweep \cite{HirFye86} grows as $1/T^3$, and the scaling with fixed fractional error is likely to be worse. Thus the present algorithm should out-perform Hirsch and Fye for low temperatures. 
(2) There is no imaginary time discretization error in the present approach. 
(3) A big advantage of our spin-chain cluster algorithm is that it is essentially unaffected by the value of $U$. In contrast, the relevant dimensionless parameter determining the computational effort for the Hirsch and Fye Anderson-model algorithm is $U/T$.  Since $U$ is proportional to the effective bandwidth $D_{\rm eff}$ within which charge fluctuation can be neglected, the ratio $D_{\rm  eff}/T$ is limited to a relatively modest value ($\sim\! 100$) in the Hirsch and Fye algorithm. This
value is not large enough to fully study the implications of the mesoscopic 
fluctuations on the renormalization of the coupling constant which 
dominates Kondo physics \cite{Kaul05}.
(4) We wish to stress that although we have illustrated our spin-chain approach for the two simplest impurity models, the general method is extremely flexible and is applicable to a wide class of multi-band fermionic or spin quantum impurity problems. The flexibility is already illustrated by the fact that the implementation for the Kondo model Eq.~(\ref{eq:Hi2}) is essentially as simple as for the Anderson model Eq.~(\ref{eq:Hi1}). The Kondo version \cite{FyeHir89} of the Hirsch and Fye algorithm is, in contrast, substantially more complex than its original Anderson counterpart, and appears in any case significantly less used. Furthermore, although in Hirsch and Fye, the 1-channel problem may not suffer from the sign problem, in general the 2-channel problem does \cite{JarPan}. In our method there is no sign problem in either case. That the $N$-channel Kondo problem \cite{Noz80} does not suffer from a sign problem in our method can be easily proved because, first, fermions in one channel cannot permute, and, second, since channel number is conserved, permutation of fermions in distinct channels is also forbidden.  
Finally, based on our experience we believe that the spin-chain algorithm may be extendable
to more complex impurity problems, involving for instance more than one orbital. 

To summarize, we have proposed a QMC method that can be formulated for a large class of quantum impurity problems in continuous time without a sign problem. We have illustrated this method specifically to simulate the Anderson and Kondo single-impurity models, and shown how it is particularly useful for simulations of these models in the context of mesoscopic physics. The effort in computing quantities grows as a function of $T_{\rm K} N/ T$.  In this work $N$ of order $5000$ and $T_{\rm K}/T$ of the order of $10$ could easily be reached. If significantly larger values of $N$ are required, it is possible to approximate the problem using logarithmic blocking, similar to that used in the NRG approach. This reduces the effective number of levels in the spin-chain and should allow one
to probe significantly lower temperature. The ideas presented here can be extended easily to the multi-channel Kondo model \cite{Noz80}, and the algorithm should be applicable to DMFT calculations \cite{GeoKot}.

This work was supported in part by the NSF (DMR-0103003).

\end{document}